\documentclass{elsart}
\usepackage{graphics}
\begin{document}
\begin{frontmatter}
\title{Applying the flow equations to QCD}
\author{Hans-Christian Pauli} 
\address{Max-Planck-Institut f\"ur Kernphysik, 
         D-69029 Heidelberg, Germany}
%
%  \date{23 September 2000}  
%
\begin{abstract}
  The effective $q\bar q$-interaction 
  is derived from Lagrangian QCD in the front form
  by means of the flow equations.
  It coincides with previous results. 
\end{abstract}
\hyphenation{ }
\end{frontmatter}

Before plunging into the paraphernalia of field theoretical details, 
Franz Wegner might allow me to outline his method \cite{Weg94}  
in yet an other language. 
Working with a Hamiltonian in an explicit representation,
it is always possible to divide the complete Hilbert space 
arbitrarily into two pieces, the $P$- and the $Q$-space.
Suppose, one is unable to solve the eigenvalue problem, 
because of say computer limitations. 
The method of flow equations allows then to unitarily transform 
the Hamiltonian matrix analytically into a 
block-diagonal effective Hamiltonian,
{\it i.e.} 
\begin{equation}
  H =\pmatrix{PHP & PHQ \cr QHP & QHQ\cr}
  \quad\longrightarrow\quad
  H_{\rm eff} =\pmatrix{PH_{\rm eff}P & 0 \cr 0 & QH_{\rm eff}Q\cr}
,\label{eq:i3}\end{equation}
where $P$ and $Q=1-P$ are the respective projection operators.
(The methods also works for Hamiltonians with more
than 2$\times$2 blocks.) 
The `reducible' matrix can then be diagonalized separately 
for either of the two spaces. 
The transformed Hamiltonian is a function of the
continuous parameter $0\leq l\leq\infty$, 
\begin{equation}
   \frac{d}{dl} H(l) = [\eta(l), H(l)]
   ,\hskip4em
   \eta(l) = [H_d(l), H(l)]
.\label{eq:i4}\end{equation}
The generator of the transformation $\eta(l)$ 
is subject to some choice \cite{Weg00}.
The Hamiltonian is separated conveniently into a block-diagonal
part $H_d(l)$ and into the rest $H_r(l)\equiv H(l)-H_d(l)$ 
which is purely off-diagonal: 
\begin{equation}
  H_d(l) =\pmatrix{PH(l)P & 0 \cr 0 & QH(l)Q \cr}
  ,\hskip2em
  H_r(l) =\pmatrix{0 & PH(l)Q \cr QH(l)P & 0 \cr}
.\end{equation}
In many cases of practical interest one can interpret
this rest as a `residual interaction'. If the rest vanishes,
or if it is exponentially small, one has solved an important
part of the problem. 
The generator is always off-diagonal:
$P\eta(l)P=Q\eta(l)Q=0$.
The flow equations (\ref{eq:i4}) for the diagonal and the 
off-diagonal block can then be divided into two coupled equations,
\begin{equation}
   \frac{d}{dl} (PHP) = P\eta QHP - PHQ\eta P
  ,\ %
   \frac{d}{dl} (PHQ) = P\eta QHQ - PHP\eta Q
,\label{eq:i9}\end{equation}
and into the trivial identity $P\eta Q = PHPHQ - PHQHQ$.
The corresponding equation for $QHQ$  will not be needed.

A possible measure for the `block-off-diagonality' is 
${\mathcal O}\equiv {\rm Tr} PHQHP\geq0$.
Its derivative is determined by the flow equations,
\begin{eqnarray}
    \frac{d}{dl} {\mathcal O}
    &=& {\rm Tr} \Big( P\eta Q (QHQHP - QHPHP)\Big)
\label{eq:i11}\\
    &+&  {\rm Tr} \Big((PHPHQ - PHQHQ) Q\eta P \Big) 
     = 2 {\rm Tr} (P\eta Q\eta P)\leq 0
,\nonumber\end{eqnarray}
because $\eta$ is anti-hermitean.   
Since ${\mathcal O}$ is positive with a negative slope,
the generator of Eq.(\ref{eq:i4}) ensures a monotonic 
decrease: The off-diagonal block tends to vanish 
in the limit $l\rightarrow\infty$.
With increasing $l$
the effective Hamiltonian $H_{\rm eff}$ becomes
`more and more block-diagonal'. 

Almost simultaneously with \cite{Weg94}  
and with a similar objective, 
Glazek and Wilson \cite{GlaWil93} have proposed 
the `similarity transform'
to render a Hamiltonian `more and more diagonal'.

One should note the subtle but important difference
between the two methods. 
Glazek and Wilson address
to generate a band matrix whose width $\lambda$ is decreased
by analytical unitary transformations.
In the asymptotic limit $\lambda\sim 1/l \rightarrow 0$
the Hamiltonian becomes diagonal.
It is asked for too much, perhaps, to find a 
transform which diagonalizes a Hamiltonian completely. 
Flow equations have been applied to many problems,
including solid-state physics, 
as well as field and gauge field theory.
I do not even attempt to review all the work but refer to 
Wegner's talk \cite{Weg00}. 
I have taken great inspiration from his work with
P.~Lenz \cite{LenWeg96}.

\section{Applying the flow equations to quantum chromo-dynamics}
\label{sec:b}

In the next two sections I apply the flow equations 
to Non-Abelian gauge theory in the front form
along the lines of earlier unpublished work \cite{GubWegPau98}.
The thinking in terms of block matrices is almost
ideally suited  for gauge theories,
where the Fock-space expansions give quite naturally
a block structure of the Hamiltonian, 
{\it cf.} for example Fig.~2 in \cite{BroPauPin98}.
It is thus reasonable to identify the $P$-space
with the lowest Fock-space sector, the one with one quark
and one anti-quark ($q\bar q$). 
The $Q$-space is then literally the `rest'.
In front form, the Hamiltonian proper is  $H=P^-$, with
\begin{equation}
   P^-=T+V+F+S
.\end{equation}
This as well as all other formulas in this section 
are taken without further notice
from the Compendium \cite{Com00}, using  
the same notation and conventions as there.
The {\em kinetic energy} $T$ is the only piece of $P^-$
which survives the limit $g\rightarrow 0$.
The coupling constant $g$ is related to the
(strong) fine structure constant $\alpha_s=g^2/(4\pi\hbar c)$.
The {\em vertex interaction} $V$ is the relativistic
interaction {\it per se}. It is linear in $g$ and
changes the particle number.
The {\em instantaneous interactions} $F$ and $S$ 
(the `forks' and the `seagulls'), are quadratic in $g$
and are artifacts of the light-cone gauge.
Both may change the particle number.~--
As a technical trick and for gaining more transparency 
the instantaneous interactions are first omitted. 
They will be re-installed at the end of the calculation.
By the same reason all color factors will be suppressed first 
and re-installed at the end.
The Hamiltonian is then
\begin{equation}
   H = T + V
.\end{equation}
Since the front-form $V$ has non-vanishing matrix elements
between Fock states only, if the particle number differs by 
\underline{1 and only 1}, 
the initial $P$-space Hamiltonian is diagonal, $PHP=T$, see below.
As opposed to that the effective Hamiltonian in the $P$-space
has an effective interaction $U$,
\begin{equation}
   PH_{\rm eff}P = T + U
.\label{peq:8}\end{equation}
The `residual interaction' is then just the vertex interaction
$H_r\equiv V$.
The $Q$-space Hamiltonian remains a complicated
operator for all $l$. It will not be addressed here.
Objective of the present work is the calculation of $U$.

The way the flow equations work is
that $V(l)$ de-creases while $U(l)$ in-creases for growing $l$. 
As emphasized by Wegner \cite{Weg00}, flow equations
generate a whole series of many-body interactions.
They can be classified with the power of the 
coupling constant $g$.
For the reason of demonstrating the method,
I restrict here and below to the lowest non-trivial order
in $g$ without further mentioning.

The flow equations (\ref{eq:i9}) are re-ordered and re-labeled as  
\begin{equation}
   \eta (l) = [ T(l), V(l)]
,\hskip1em
   \frac{d V(l)}{dl} = [ \eta(l), T(l)]
,\hskip1em
   \frac{d U(l)}{dl} = [ \eta(l), V(l)]
.\label{eq:ii44}\end{equation}
The relativistic kinetic energy is diagonal for all $l$:
\begin{equation}
    T(l) = \int [d^3q] \left( 
   \frac{\overline m^2(p;l)+\vec{p}_{\!\perp}^{\ 2}}{p^+}\right)_q   
   \left(b_q^\dagger b_q
   +d_q^\dagger d_q 
   +a_q^\dagger a_q\right)
.\end{equation}
The conventions in the Compendium include that
$[d^3q]$ stands for the integration over the longitudinal momentum
$p^+$ and the transversal momenta $\vec{p}_{\!\perp}$,
as well as the summation over helicity, color (glue) and flavor.
The single particle energies depend potentially 
on the flow parameter through
the effective mass $\overline m^2(p;l)$.
This dependence will be suppressed below, however, 
since it comes in higher order of $g$.
Hence forward, $p^+$ and $p_{\!\perp}$ will be collected in the 
3-vector $\vec p=(p^+,\vec p_{\!\perp})$.
The interaction term is 
\begin{eqnarray}
    V(l) &=& {1\over\sqrt{16\pi^3}} 
   \int\!{[d^3q_1]\over\sqrt{p ^+_1}} 
   \int\!{[d^3q_2]\over\sqrt{p ^+_2}} 
   \int\!{[d^3q_3]\over\sqrt{p ^+_3}} 
   \ \delta^{(3)}(\vec p_1 - \vec p_2 - \vec p_3) 
   \,\overline g(\vec p_1,\vec p_2,\vec p_3;l)
\nonumber\\ &\times&  
   \left[ b^\dagger_1 b_2 a_3 \, 
   (\overline u_1 \slash\!\!\!\epsilon_3 u_2) 
   -d_1^\dagger d _2 a_3 \, 
   (\overline v_2 \slash\!\!\!\epsilon_3 v_1)
   +a_1^\dagger d_2 b _3 \, 
   (\overline v_2 \slash\!\!\!\epsilon_1^\star u_3) 
   \right] + h.c.
.\label{eq:25}\end{eqnarray}
The three-gluon vertex is omitted here because it does not 
contribute to the effective interaction 
to the lowest non-trivial order.
The effective coupling 
$\overline g(\vec p_1,\vec p_2,\vec p_3;l)$,
with the initial value 
$\overline g(\vec p_1,\vec p_2,\vec p_3;l=0)= g$,
contains all the $l$-dependence. 
The generator  
$ \eta = [ T, V]$ becomes 
\begin{eqnarray}
    \eta (l) &=& {1\over\sqrt{16\pi^3}}
   \int\!{[d^3q_1]\over\sqrt{p ^+_1}} 
   \int\!{[d^3q_2]\over\sqrt{p ^+_2}} 
   \int\!{[d^3q_3]\over\sqrt{p ^+_3}}
   \,\delta^{(3)}(\vec p_1 - \vec p_2 - \vec p_3) 
   \,\overline\eta (\vec p_1,\vec p_2,\vec p_3;l)
\nonumber\\
   &\times&  
   \left( b^\dagger_1 b_2 a_3 \, 
   [\overline u_1 \slash\!\!\!\epsilon_3 u_2] 
   -d_1^\dagger d _2 a_3 \, 
   [\overline v_2 \slash\!\!\!\epsilon_3 v_1]
   +a_1^\dagger d_2 b _3 \, 
   [\overline v_2 \slash\!\!\!\epsilon_1^\star u_3] 
   \right) - h.c.
.\label{eq:28}\end{eqnarray}
The structure of $ \eta$ is very similar to $ V$, 
because $ T$ is diagonal, thus
\begin{equation}
   \overline\eta (\vec p_1,\vec p_2,\vec p_3;l) = 
   \overline g(\vec p_1,\vec p_2,\vec p_3;l)
   c(\vec p_1,\vec p_2,\vec p_3;l)
.\end{equation}
The commutator with $T$ generates the `commutator function'
\begin{equation}
   c(\vec p_1,\vec p_2,\vec p_3;l) \equiv 
   e(\vec p_1;l) - e(\vec p_2;l) - e(\vec p_3;l)
,\end{equation}
with the single particle energies  
$e(\vec p;l) = {(\overline m^2(p;l)+\vec{p}_{\!\perp}^{\ 2})}/{p^+}$.
Symmetries like 
$\overline g(p^{\:\prime}_1,p_1,p^{\:\prime}_1-p_1;l)=
 \overline g(p_1,p^{\:\prime}_1,p_1-p^{\:\prime}_1;l)$
are so obvious that they, as well as
$\overline \eta(p^{\:\prime}_1,p_1,p^{\:\prime}_1-p_1;l)=-
 \overline\eta(p_1,p^{\:\prime}_1,p_1-p^{\:\prime}_1;l)$, 
will not be mentioned hence forward.
Similarly, the derivative 
$d V (l)/dl = [ \eta(l), T(l)]$
becomes
\begin{eqnarray}
   \frac{d V (l)}{dl}  = 
   &-& 
   \frac{1}{\sqrt{16\pi^3}}
   \int\!{[d^3q_1]\over\sqrt{p ^+_1}} 
   \int\!{[d^3q_2]\over\sqrt{p ^+_2}}  
   \int\!{[d^3q_3]\over\sqrt{p ^+_2}} 
   \,\delta^{(3)}(\vec p_1 - \vec p_2 - \vec p_3)
\\
   &\times& 
   \overline \eta(\vec p_1,\vec p_2,\vec p_3;l)
   c(\vec p_1,\vec p_2,\vec p_3;l) 
\nonumber\\
   &\times& 
   \left( b^\dagger_1 b_2 a_3 \, 
   [\overline u_1 \slash\!\!\!\epsilon_3 u_2] 
   -d_1^\dagger d _2 a_3 \, 
   [\overline v_2 \slash\!\!\!\epsilon_3 v_1]
   +a_1^\dagger d_2 b _3 \, 
   [\overline v_2 \slash\!\!\!\epsilon_1^\star u_3] 
   \right) 
   + h.c.
.\nonumber\end{eqnarray}
Finally, the derivative   
${d U(l)}/{dl} = [\eta(l),V(l)]$ becomes
\begin{eqnarray}
   &&\frac{d  U (l)}{dl} = -{1\over16\pi^3} 
   \int\!{[d^3q_1]\over\sqrt{p ^+_1}}
   \int\!{[d^3q_2]\over\sqrt{p ^+_2}} 
   \int\!{[d^3q_3]\over\sqrt{p ^+_3}}
   \int\!{[d^3q_1^\prime]\over\sqrt{p ^{\prime+}_1}} 
   \int\!{[d^3q_2^\prime]\over\sqrt{p ^{\prime+}_2}} 
   \int\!{[d^3q_3^\prime]\over\sqrt{p ^{\prime+}_3}} 
\nonumber\\ 
   &&\bigg(
   \delta^{(3)}(\vec p_1 - \vec p^{\:\prime}_1 - \vec p_3) 
   \delta^{(3)}(\vec p^{\:\prime}_2 - \vec p_2 - \vec p^{\:\prime}_3) 
   \left[b_1^\dagger b_{1'} a_3,
   a_{3'}^\dagger d_{2}^\dagger d _{2'}\right]
   [\overline u_1\slash\!\!\!\epsilon _3 u _{1'}]
   [\overline v_{2'}
   \slash\!\!\!\epsilon^\star_{3'} v_{2}]  
\nonumber\\ 
   &&\phantom{aaa}\times\Big(
   \overline\eta(\vec p_1,\vec p^{\:\prime}_1,\vec p_3;l) 
   \overline   g(\vec p^{\:\prime}_2,\vec p_2,\vec p^{\:\prime}_3;l) + 
   \overline\eta(\vec p^{\:\prime}_2,\vec p_2,\vec p^{\:\prime}_3;l) 
   \overline   g(\vec p_1,\vec p^{\:\prime}_1,\vec p_3;l) \Big) +
\nonumber\\ 
   &&\phantom{a} 
    \delta^{(3)}(\vec p^{\:\prime}_1 - \vec p_1 - \vec p_3) 
    \delta^{(3)}(\vec p_2 - \vec p^{\:\prime}_2 - \vec p^{\:\prime}_3) 
   \left[d_{2}^\dagger d _{2'} a_{3'} ,
   a_{3}^\dagger b_{1}^\dagger b_{1'} \right]
   [\overline u_{1}\slash\!\!\!\epsilon_{3}^\star u_{1'}] 
   [\overline v_{2'}\slash\!\!\!\epsilon_{3'} v_{2}] 
\nonumber\\  
   &&\phantom{aaa}\times\Big( 
   \overline\eta(\vec p^{\:\prime}_1,\vec p_1,\vec p_3;l)\,
   \overline   g(\vec p_2,\vec p^{\:\prime}_2,\vec p^{\:\prime}_3;l)+ 
   \overline\eta(\vec p_2,\vec p^{\:\prime}_2,\vec p^{\:\prime}_3;l)\,
   \overline   g(\vec p^{\:\prime}_1,\vec p_1,\vec p_3;l)
   \Big) \bigg) 
\label{eq:32}\end{eqnarray} 
plus terms containing commutators like 
$[a^\dagger db,b^\dagger d^\dagger a]$  or 
$[a^\dagger db,a^\dagger a^\dagger a]$   
contributing only in higher order of $g$.
After execution of the various commutators it is obvious 
that $U$ has the general form \ ({\it cf.} Eq.(\ref{peq:8})\ )
\begin{eqnarray}
     U(l) = 
   \int\![d^3q_1] \int\! [d^3q_2] \int\! [d^3q'_1] \int\! [d^3q'_2]
   &{}&\delta^{(3)}(\vec p_1-\vec p^{\:\prime}_1
                   +\vec p_2-\vec p^{\:\prime}_2)
\nonumber\\ &\times& 
   b^\dagger _{q_1} b _{q'_1} \,d^\dagger _{q_2} d _{q'_2} 
   \ \overline U(q_1,q_2;q'_1,q'_2;l)
.\label{eq:33}\end{eqnarray}
It suffices thus to calculate the derivative of a $c$-number
\begin{eqnarray}
   &&{d\overline U \over dl}  = - {1\over 16\pi^3}
   \left( \theta(p_1^+ - p^{\prime+}_1) + \theta(p^{\prime+}_1 - p_1^+) 
   \right)
\nonumber\\ &\times& \Big(
   \overline\eta(\vec p^{\:\prime}_1,\vec p_1,\vec p^{\:\prime}_1-\vec p_1;l) 
   \overline g  (\vec p_2,\vec p^{\:\prime}_2,\vec p_2-\vec p^{\:\prime}_2;l)
\nonumber\\   &&\phantom{mmmmmmmmbbbaaaaaaa}+ 
   \overline g(\vec p^{\:\prime}_1,\vec p_1,
   \vec p^{\:\prime}_1-\vec p_1;l) 
   \overline \eta(\vec p_2,\vec p^{\:\prime}_2,
   \vec p_2-\vec p^{\:\prime}_2;l) \Big)
\nonumber\\ &\times& 
   {[\overline u(p_1,\lambda_1)\gamma^\mu u(p^{\:\prime}_1,\lambda'_1)]
   \over\sqrt{p _1^{+}}\phantom{\gamma^\mu}\sqrt{p ^{\prime+}_1}}
   \frac{d_{\mu\nu}(q)}{q^+} 
   \frac
   {[\overline v(p^{\:\prime}_2,\lambda'_2)\gamma^{\nu} v(p_2,\lambda_2)]}
   {\sqrt{p ^{\prime+}_2}\phantom{\gamma^\mu}\sqrt{p _2^+} }   
.\label{eq:ii58}\end{eqnarray} 
The polarization tensor $d_{\mu\nu}$ arises by the summing
over the helicities, {\it i.e.}
$d_{\mu\nu}(q)\equiv \sum_{\lambda}
 \epsilon_{\mu}(q,\lambda)\epsilon^{*}_{\nu}(q,\lambda)$.
The 4-momentum of the photon is denoted by $q^\mu$. 
The step function is $\theta(x)=1$ for $x\ge 0$, 
thus $\theta(x)+\theta(-x)=1$.

\section{Integrating the flow equations and shaping the notation}
\label{sec:c}

Eq.(\ref{eq:ii44}) gives the first order flow equations  
in operator form. 
After evaluating
the matrix elements they reduce to two coupled algebraic equations: 
\begin{eqnarray}
   \frac{d}{dl} \overline g(\vec p_1,\vec p_2,\vec p_3;l) &=& -
   c(\vec p_1,\vec p_2,\vec p_3;0)\ %
   \overline\eta(\vec p_1,\vec p_2,\vec p_3;l)
,\label{eq:37}\\ 
   \overline \eta(\vec p_1,\vec p_2,\vec p_3;l) &=& \phantom{-}
   c(\vec p_1,\vec p_2,\vec p_3;0)\ %
   \overline g(\vec p_1,\vec p_2,\vec p_3;l)
.\label{eq:38}\end{eqnarray}
They can be combined into a single one
\begin{equation}
   \frac{d}{dl} \overline g(\vec p_1,\vec p_2,\vec p_3;l) = -
             c^2(\vec p_1,\vec p_2,\vec p_3;0)
   \ \overline g(\vec p_1,\vec p_2,\vec p_3;l)
,\end{equation}
and integrated with the possible solution 
\begin{equation}
   \overline g(\vec p_1,\vec p_2,\vec p_3;l) = 
   g\ {\rm exp} \left(-l\ c^2(\vec p_1,\vec p_2,\vec p_3;0)\right)
.\label{eq:f67}\end{equation}
Since the $p'$s appear in combinations like 
$(\vec p^{\:\prime}_1,\vec p_1,\vec p^{\:\prime}_1-\vec p_1;l)$
(see {\it f.e.} Eq.(\ref{eq:ii58})\,), 
I introduce conveniently the energy differences 
\begin{eqnarray}
   w_{q}      &\equiv& -
   c(\vec p^{\:\prime}_1,\vec p_1,\vec p^{\:\prime}_1-\vec p_1;0) =
   (p^{\:\prime}_1-p_1)^- - p^{\:\prime-}_1 + p_1^- 
,\\
   w_{\bar q} &\equiv& -
   c(\vec p_2,\vec p^{\:\prime}_2,\vec p_2-\vec p^{\:\prime}_2;0) =
   (p_2-p^{\:\prime}_2)^- - p_2^- + p^{\:\prime-}_2  
.\label{eq:40a} \end{eqnarray}
Typically, they occur along the quark ($w_{q}$)
and the anti-quark line $(w_{\bar q})$.
They are in simple relationship 
to both the 4-momentum of the exchanged photon
\begin{equation}
   q_\mu = p^{\:\prime}_{1\mu}-p_{1\mu}+\eta_\mu \frac{w_{q}}{2}
         = p_{2\mu}-p^{\:\prime}_{2\mu}+\eta_\mu \frac{w_{\bar q}}{2}
\,,\label{eq:ii66}\end{equation}
and to the Feynman 4-momentum transfers along the two lines 
\begin{equation}
   Q_{q}^2 \equiv -(p^{\:\prime}_1-p_1)^2 = q^+w_{q} 
   ,\hskip4em
   Q_{\bar q}^2 \equiv -(p_2-p^{\:\prime}_2)^2 = q^+w_{\bar q}
.\label{eq:w25}\end{equation}
The latter need not be equal in a Hamiltonian approach.
The null 4-vector $\eta^{\mu}$ is defined by $\eta^{\mu}=(0,\vec{0},2)$
and cannot be confused with the operator $\eta$.
The Feynman\--momentum transfer $Q^2$ is always positive 
and a more physical quantity than an energy difference. 
The $w$'s will be substituted therefore by the $Q$'s 
as long as no misunderstanding can arise. 
In fact, I shall use often the 
{\em mean-square momentum transfer}
\begin{equation}
   Q^2 \equiv 
   \frac12 (Q_{q}^2+Q_{\bar q}^2) = 
   \frac{q^+}{2}(w_{q}+w_{\bar q})
,\label{eq:r15}\end{equation}
and the {\em mean-square difference}
\begin{equation}
  \delta Q^2 \equiv 
  \frac12 (Q_{q}^2-Q_{\bar q}^2) =  
  \frac{q^+}{2}(w_{q}-w_{\bar q})
.\end{equation}
Also the polarization tensor $d_{\mu\nu}$ 
in Eq.(\ref{eq:ii58}) can be simplified with these definitions.
Since 
$d_{\mu\nu}(q)=-g_{\mu\nu}+{(\eta_{\mu}q_{\nu}+\eta_{\nu}q_{\mu})}/{q^+}$ 
appears always in combinations with the spinors, 
the Dirac equation 
$(p_1-p^{\:\prime}_1)_\mu \overline u(p_1)\gamma^\mu u(p^{\:\prime}_1) = 0$
can be used to write
\begin{equation}
   q_\mu \overline u(p_1,\lambda_1)\gamma^\mu u(p^{\:\prime}_1,\lambda'_1) 
   = {w_{q}\over 2}
   \eta_\mu \,\overline u(p_1,\lambda_1)\gamma^\mu u(p^{\:\prime}_1,\lambda'_1)
.\end{equation} 
In the context of Eq.(\ref{eq:ii58})
one can use thus   
$d_{\mu\nu}(q)=-g_{\mu\nu}+\eta_\mu\eta_\nu Q^2/{(q^+)}^2$. 

Following Glazek and Wilson \cite{GlaWil93}, 
I introduce a dimensionless 
similarity function $f(l)$ (with 
boundary conditions $f(0)\equiv 1$ and $f(\infty)\equiv 0$) by
\begin{equation}
   f(\vec p_1,\vec p_2,\vec p_3;l) \equiv \frac{1}{g}
   \overline g(\vec p_1,\vec p_2,\vec p_3;l) 
.\end{equation}
Note that $f$ describes the decay rate of the off-diagonal 
vertex interaction $V$.
The rate of decay, however, is not a unique function,
as will be seen below.
In line with recent developments \cite{Weg00},
I want to keep $f=f(w;l)$ as a general function,
particularly since $\overline \eta$ can be rewritten as 
\begin{equation}
   \overline \eta(l) = - \frac{1}{w}
   \left(\frac{d{\rm ln}f(w;l)}{dl}\right)\overline g(l)
.\label{eq:55a}\end{equation}
Finally, I define  
\begin{equation}
   N_{q} \equiv -\int_{0}^\infty\!dl' 
   \frac{d f (w_{q};l')}{dl'} f (w_{\bar q};l')
,\hskip2em
   N_{\bar q} \equiv 1 - N_{q} 
,\label{eq:ii77}\end{equation}
which are kind of dimensionless occupation numbers.

With these definitions I return to the problem of
integrating Eq.(\ref{eq:ii58}). 
Its relevant part is
\begin{eqnarray}
   && 
   \overline\eta(\vec p^{\:\prime}_1,\vec p_1,\vec p^{\:\prime}_1-\vec p_1;l)
   \overline g  (\vec p_2,\vec p^{\:\prime}_2,\vec p_2-\vec p^{\:\prime}_2;l) +
   \overline\eta(\vec p_2,\vec p^{\:\prime}_2,\vec q;l)
   \overline g  (\vec p^{\:\prime}_1,\vec p_1,\vec q;l) 
\nonumber\\
   &&\phantom{mmmmm}= - g^2 \left(
   \frac{1}{w_{q}}
   \frac{d f (w_{q};l)}{dl} f (w_{\bar q};l) +
   \frac{1}{w_{\bar q}}\frac{d f (w_{\bar q};l)}{dl} 
   f (w_{q};l)\right)
.\label{eq:ii72}\end{eqnarray} 
The $l$-integration leads to
\begin{eqnarray}
   &&\int_0^\infty\!dl'\Big(
   \overline\eta(\vec p^{\:\prime}_1,\vec p_1,\vec p^{\:\prime}_1-\vec p_1;l')
   \overline   g(\vec p_2,\vec p^{\:\prime}_2,\vec p_2-\vec p^{\:\prime}_2;l')
 + \overline\eta(\vec p_2,\vec p^{\:\prime}_2,\vec q;l')
   \overline   g(\vec p^{\:\prime}_1,\vec p_1,\vec q;l') \Big)
\nonumber\\  
   &&\phantom{mmmmm}= g^2 
   \left(\frac{N_{q}}{w_{q}} +\frac{N_{\bar q}} {w_{\bar q}}\right) = 
   g^2 q^+ \left(
   \frac {N_{q}}  {Q_{q}^2} + \frac {N_{\bar q}} {Q_{\bar q}^2}\right) 
.\end{eqnarray} 
It is natural to assume \cite{Weg00} that the similarity function 
$f(w;l)$ is a homogeneous function of its arguments
$f(w;l)=f(w^\kappa l)$, with some exponent $\kappa$. 
Two types of similarity functions are considered:
\begin{eqnarray}
   \mbox{Gaussian\phantom{iial} cut-off ($\kappa=2$): }
   &&f(w;l)=\exp(-w^2 l), \phantom{a} 
   N_{q}=\frac{w_{q}^2}{w_{q}^2+w_{\bar q}^2}
,\label{eq:w32}\\
   \mbox{Exponential cut-off ($\kappa=1$): }
   &&f(w;l)=\exp(-w l), \phantom{aa}  
   N_{q}=\frac{w_{q}}{w_{q}+w_{\bar q}}
.\label{peq:36}\end{eqnarray}
The first case corresponds to Eq.(\ref{eq:f67}).
The second shows that the effective interaction depends 
explicitly on the similarity function. 
The requirement of block diagonalizing the Hamiltonian 
determines the generator only up to a 
unitary transformation \cite{Weg00}.

\section{The effective $q\bar q$-interaction from Hamiltonian flow}
\label{sec:d}

Collecting terms, the effective interaction 
from Eq.(\ref{eq:ii58}) becomes 
\begin{eqnarray*}
   \overline U = g^2 \frac {d_{\mu\nu}(q)} {16\pi^3}
   \frac {[\overline u(p_1,\lambda_1)\gamma^\mu
       u(p^{\:\prime}_1,\lambda'_1)]} 
       {\sqrt{p _1^+}\phantom{\gamma^\mu}\sqrt{p ^{\prime+}_1}}
   \frac {[\overline v(p^{\:\prime}_2,\lambda'_2)\gamma^\nu
    v(p_2,\lambda_2))]} 
    {\sqrt{p ^{\prime+}_2}\phantom{\gamma^\mu}\sqrt{p _2^+}}
   \left(\frac{N_{q}}      {Q_{q}^2} +
         \frac{N_{\bar q}} {Q_{\bar q}^2}\right)
.\end{eqnarray*}
It is easy to restore the color factors.
They sit only in the vertex interactions and go through 
untouched by the flow equations, thus
\begin{eqnarray*}
   \lefteqn{
   \overline U = g^2 \frac {d_{\mu\nu}(q)} {16\pi^3}
   \frac {[\overline u(p_1,\lambda_1)T^a\gamma^\mu
       u(p^{\:\prime}_1,\lambda'_1)]} 
       {\sqrt{p _1^+}\phantom{\gamma^\mu}\sqrt{p ^{\prime+}_1}}
   \frac {[\overline v(p^{\:\prime}_2,\lambda'_2)T^a\gamma^\nu
    v(p_2,\lambda_2))]} 
    {\sqrt{p ^{\prime+}_2}\phantom{\gamma^\mu}\sqrt{p _2^+}}
   \left(\frac{N_{q}}      {Q_{q}^2} +
         \frac{N_{\bar q}} {Q_{\bar q}^2}\right)
.}\end{eqnarray*}
If the $d_{\mu\nu}(q)$ is substituted, this turns into  
\begin{eqnarray}
   \overline U &=& -\frac {g^2} {16\pi^3}
   \frac {[\overline u(p_1,\lambda_1)T^a\gamma^\mu
       u(p^{\:\prime}_1,\lambda'_1)]} 
       {\sqrt{p _1^+}\phantom{\gamma^\mu}\sqrt{p ^{\prime+}_1}}
   \frac {[\overline v(p^{\:\prime}_2,\lambda'_2)T^a\gamma^\nu
    v(p_2,\lambda_2)]} 
    {\sqrt{p ^{\prime+}_2}\phantom{\gamma^\mu}\sqrt{p _2^+}}
\nonumber\\
   &&\phantom{mmmmmmmmm}\times 
   \left(g_{\mu\nu}-\eta_\mu \eta_\nu\frac {Q^2}{(q^+)^2}\right)
   \left(\frac{N_{q}} {Q_{q}^2} +
         \frac{N_{\bar q}} {Q_{\bar q}^2}\right)
.\label{peq:37}\end{eqnarray}
The result holds actually for 
the general non-Abelian theory $SU(n_c)$.

It is equally straightforward to re-install
the instantaneous interaction:
\begin{equation}
   \overline U_{\rm i} =  \frac{-g^2}{16\pi^3}
   \frac{[\overline u(p_1,\lambda_1) T^a\gamma^\mu
       u(p^{\:\prime}_1,\lambda'_1)]}
       {\sqrt{p _1^+}\phantom{T^a\gamma^\mu}\sqrt{p _1^{'+}}}
   \frac{[\overline v(p^{\:\prime}_2,\lambda'_2)T^a\gamma^\nu
    v(p_2,\lambda_2)]}
    {\sqrt{p _2^{'+}}\phantom{T^a\gamma^\mu}\sqrt{p _2^+}}
   \frac{\eta_\mu \eta_\nu}{(q^+)^2}
,\label{eq:46}\end{equation}
{\it cf.} diagram $S_{3,1}$ in Table~5 of \cite{Com00}.
To the order considered here it is independent of $l$.
Adding it in by 
$\overline U\longrightarrow\overline U+\overline U_{\rm i}$ gives
\begin{eqnarray}
   \overline U = &-& \frac{g^2}{16\pi^3}
   \frac{[\overline u(p_1,\lambda_1)T^a\gamma^\mu
       u(p^{\:\prime}_1,\lambda'_1)]}
       {\sqrt{p _1^+}\phantom{T^a\gamma^\mu}\sqrt{p ^{\:\prime+}_1}}
   \frac{[\overline v(p^{\:\prime}_2,\lambda'_2)T^a\gamma^\nu
    v(p_2,\lambda_2)]}
    {\sqrt{p ^{\prime+}_2}\phantom{T^a\gamma^\mu}\sqrt{p _2^+}} 
\nonumber\\
   &&\phantom{mm}\times\left[g_{\mu\nu} 
   \left(\frac{N_{q}}{Q_{q}^2} +
         \frac{N_{\bar q}} {Q_{\bar q}^2}\right) + 
   \eta_\mu\eta_\nu
   \left(\frac{N_{q}} {Q_{q}^2} - 
         \frac{N_{\bar q}} {Q_{\bar q}^2}\right)\,
   \frac {\delta Q^2}{(q^+)^2}
   \right]
.\label{eq:47}\end{eqnarray}
Contracting the Lorentz indices 
(with $\gamma^\mu\eta_\mu = \gamma^+$) gives
\begin{eqnarray}
   \overline U =&-&   
   \frac{\alpha_s}{4\pi^2} 
   \frac{
   [\overline u(p_1,\lambda_1)T^a\gamma^\mu u(p^{\:\prime}_1,\lambda'_1)]\,
   [\overline v(p^{\:\prime}_2,\lambda'_2)T^a\gamma_\mu v(p_2,\lambda_2)]}
   {\sqrt{p _1^+\ p _2^+ \quad p _1^{\prime+}\ p _2^{\prime+} }} 
   \left(\frac{N_{q}}      {Q_{q}^2} +
         \frac{N_{\bar q}} {Q_{\bar q}^2}\right)
\nonumber\\ &-&
   \frac{\alpha_s}{4\pi^2} 
   \frac{
   [\overline u(p_1,\lambda_1)T^a\gamma^+ u(p^{\:\prime}_1,\lambda'_1)]\,
   [\overline v(p^{\:\prime}_2,\lambda'_2)T^a\gamma^+ v(p_2,\lambda_2)]}
   {\sqrt{p _1^+\ p _2^+ \quad p _1^{\prime+}\ p _2^{\prime+} }} 
\nonumber\\ &&\phantom{mmmmmmmmmmmmmmm}\times
   \frac {\delta Q^2}{(q^+)^2} 
   \left(\frac{N_{q}}      {Q_{q}^2} - 
         \frac{N_{\bar q}} {Q_{\bar q}^2}\right)
.\label{eq:56}\end{eqnarray} 
One notes the non-integrable quadratic singularity $(q^+)^2$  
appearing both in the dynamic amplitude Eq.(\ref{peq:37})
and in the instantaneous interaction Eq.(\ref{eq:46}).
In the perturbative calculation of the $q\bar q$ scattering amplitude, 
these singularities cancel each other exactly, 
see for example Sect.~3D of \cite{BroPauPin98}. 
Also in the present approach they tend to cancel,
but a residual piece remains in Eq.(\ref{eq:56}).
It is weighted with a coefficient $\delta Q^2$ which
possibly can be considered small, but beyond that,
the term is dangerous because it violates manifestly 
gauge invariance (reflected in the matrix elements 
$\langle\gamma^+\gamma^+\rangle$).
The quadratic singularity carries however also 
weighting factors which depend on the similarity function,
see Eqs.(\ref{eq:w32},\ref{peq:36}).
Inserting them, I get for the exponential cut-off: 
\begin{eqnarray}
   N_{q}=\frac{Q_{q}^2}{Q_{q}^2+Q_{\bar q}^2}
\mathbf{:}\hskip6em
   \left(\frac{N_{q}}      {Q_{q}^2} +
         \frac{N_{\bar q}} {Q_{\bar q}^2}\right) &=&
   \frac{1}{Q^2} 
,\\
   \frac {\delta Q^2}{(q^+)^2}\,
   \left(\frac{N_{q}} {Q_{q}^2} - 
         \frac{N_{\bar q}} {Q_{\bar q}^2}\right) &=& 0
,\end{eqnarray}
and for the Gaussian cut-off: 
\begin{eqnarray}
   N_{q}=\frac{(Q_{q}^2)^2}{(Q_{q}^2)^2+(Q_{\bar q}^2)^2}
\mathbf{:}\hskip4em
   \left(\frac{N_{q}}      {Q_{q}^2} +
         \frac{N_{\bar q}} {Q_{\bar q}^2}\right) &=&
   \frac{Q^2}{(Q^2)^2+(\delta Q^2)^2} 
,\label{peq:43}\\
   \frac {\delta Q^2}{(q^+)^2}\,
   \left(\frac{N_{q}} {Q_{q}^2} - 
         \frac{N_{\bar q}} {Q_{\bar q}^2}\right) &=&
   \frac{\delta Q^2}{(Q^2)^2+(\delta Q^2)^2}
   \frac {\delta Q^2}{(q^+)^2}\,
.\label{peq:44}\end{eqnarray}
For $\delta Q^2 \ll Q^2$ the Lorentz contracted part 
is obviously the same for both cut-offs.
The potentially dangerous coefficient of 
$\langle\gamma^+\gamma^+\rangle$, however,
behaves drastically different:
It vanishes strictly for the exponential cut-off.
For the Gaussian cut-off it can be small,
at least it behaves now like $(\delta Q^2)^2$, 
but its precise behaviour near the important region
of the Coulomb singularity at $Q^2 \sim 0$ must
be left to future analysis.
To easen my life I will work hence-forward with the
exponential cut-off and its gauge-invariant result. 
\\
Thus far, I have studied the Hamiltonian proper $H=P^-$.
But it is often advantageous to work with the Lorentz-invariant 
`light-cone Hamiltonian' $H_{\rm LC}=P_\mu P^\mu$ \cite{BroPauPin98}.
Its eigenvalues have the dimension of an invariant mass-squared.
In the intrinsic frame ($\vec P_{\!\perp}=0$) one has 
$H_{\rm LC}=P^+P^-$.
The effective interaction of $H_{\rm LC}$,
is obtained simply by
$\overline U \Longrightarrow \overline U P^+ $, thus 
\begin{equation}
   \overline U = -    
   \frac{C_c}{4\pi^2}\: 
   \frac{
   [\overline u(p_1,\lambda_1)\gamma^\mu u(p^{\:\prime}_1,\lambda'_1)]\,
   [\overline v(p^{\:\prime}_2,\lambda'_2)\gamma_\mu v(p_2,\lambda_2)]}
   {\sqrt{x(1-x)x'(1-x')}} 
   \:\frac{\alpha_s}{Q^2}\ \frac{1}{P^+}
,\label{peq:42}\end{equation} 
where $x = p_1^+/P^+$ is the longitudinal 
momentum fraction of the quark.
The color factor is 
$C_c = \sum_{a} T^a_{c_1,c_1'} T^a_{c_2',c_2}$.  
This finishes my objective:
the so obtained effective interaction for QCD
coincides identically with the result of lowest order obtained 
with the method of iterated resolvents \cite{Pau98}.
\\% \noindent
\textbf{Acknowledgement.}
I thank Franz Wegner for many illuminating discussions. 
I thank him also for our joint effort 
to shift the chaff from the wheat.
%\begin{equation}p^{\prime+}_2,\quad\vec p^{\:\prime}_2.\end{equation}
 
\end{document}